\documentclass[prd,aps,showpacs,epsf,floats]{revtex4}
\usepackage{amssymb}
\usepackage{amsfonts}
\usepackage{amsmath}

\input{tcilatex}
\begin{document}

\title{Analytical Estimate of Atmospheric Newtonian Noise Generated by
Acoustic and Turbulent Phenomena in Laser-Interferometric Gravitational
Waves Detectors}
\author{C. Cafaro$^{1}$ and S. A. Ali$^{2}$}
\affiliation{$^{1}$Dipartimento di Fisica, Universit\`{a} di Camerino, I-62032 Camerino,
Italy\\
$^{2}$Department of Physics, State University of New York at Albany, 1400
Washington Avenue, Albany, NY 12222, USA}

\begin{abstract}
We present\textbf{\ }a theoretical estimate of the atmospheric Newtonian
noise due to fluctuations of atmospheric mass densities generated by
acoustic and turbulent phenomena and we determine the relevance of such
noise in the laser-interferometric detection of gravitational waves. First,
we consider the gravitational coupling of interferometer test-masses to
fluctuations of atmospheric density due to the propagation of sound waves in
a semispace occupied by an ideal fluid delimited by an infinitely rigid
plane. We present an analytical expression of the spectrum of acceleration
fluctuations of the test-masses of the interferometer in terms of the
experimentally obtainable spectrum of pressure fluctuations. Second, we
consider the gravitational coupling of interferometer test-masses to
fluctuations of atmospheric density due to the propagation of sound waves
generated in a turbulent Lighthill process. We present an analytical
expression - in the Fourier space - of the spectrum of acceleration
fluctuations\ of the test-masses of the interferometer. Finally, we discuss
the relevance of these noise sources in the detection of gravitational waves
by comparing the estimated spectral densities of Newtonian atmospheric
noises considered here to the expected sensitivity curve of the VIRGO
detector.
\end{abstract}

\maketitle

\section{Introduction}

It is known that in the weak field approximation of Einstein's General
Relativity (GR), the linearized equations of GR are hyperbolic, thus
implying propagation of gravitational waves with the speed of light \cite%
{MTW}. The discovery by Hulse and Taylor of the binary pulsar PSR $1913-16$,
the measured orbital period of which decreases following the predictions of
GR, constitutes an indirect observational confirmation of the existence of
gravitational waves \cite{HT}.

The direct detection of gravitational waves is an important goal of
contemporary physics and modern technology allows to build ground and space
based experiments sufficiently sensitive to detect, in a direct way,
gravitational waves radiated by astrophysical objects \cite{barone}. We
emphasize that the detection of gravity waves is also important to verify
the consistency of alternative theories of gravity that are currently not
ruled out from a pure theoretical standpoint \cite{corda, nojiri}. Detection
of gravitational waves is performed by measuring the relative displacements
of several nearly free masses which carry the mirrors defining a Michelson
interferometer. The measured quantity is the time-dependent difference in
the lengths of the two orthogonal arms of the interferometer. In principle,
this form of antenna can be sensitive down to quite low frequencies. In
practice, various noise sources will limit the useful bandpass. Several
noise processes can generate spurious signals in the antenna, masking the
effect induced by the gravitational wave. Some of these processes, such as
seismic and thermal noise, induce displacements of the mirrors
("displacement noise"), while others, such as the noise induced by frequency
fluctuations of the laser, affect the phase of the optical rays even if a
real movement of the mirrors is not present ("phase noise"). In order to
define the sensitivity of the antenna it is necessary to compare the real
signal $h_{\text{GW}}\left( t\right) $ to each relevant fake signal $h_{%
\text{Noise }}\left( t\right) $. This comparison is usually expressed in
terms of the so-called "linear spectral density" which is defined as the
square root of the power spectrum of the signal \cite{barone}. In the case
of dimensionless amplitudes, the linear spectral density can be therefore
expressed in units of $\sqrt{\text{Hz}}$. Considering for instance the
French-Italian interferometric gravitational wave detector VIRGO \cite{virgo}%
, it turns out that in the low frequency range (below few tens of Hz) the
VIRGO sensitivity is limited by the thermal noise of the pendulum
suspension. Between few tens of Hz and few hundreds of Hz, the dominant
mechanism is the thermal noise of the mirrors internal modes. At higher
frequencies the VIRGO\ sensitivity curve is limited by shot-noise (noise
generated by Poisson statistical fluctuations of the number of photons in
the light beam). Another important source of noise is the so-called
gravity-gradient noise, a noise due to fluctuating Newtonian gravitational
forces that induce motions in the test masses of an interferometric
gravitational-wave detector. Gravity gradients are potentially important at
the low end of the interferometer frequency range, $f\leq 20Hz$. Another
noise source that is important at these frequencies is the vibrational
seismic noise, in which the ground's ambient motions, filtered through the
detector's vibration isolation system, produce motions of the test masses.
It should be possible and practical to isolate the test masses from these
seismic vibrations down to frequencies as low as $f\sim 3Hz$, but it does
not seem practical to achieve significant isolation from the fluctuating
gravity gradients. Thus, gravity gradients constitute an ultimate
low-frequency noise source; seismic vibrations do not. The Virgo Sensitivity
Curve is obtained by summing up in an incoherent way the spectral noise
densities of all the considered noises (seismic noise, shot noise, thermal
noise, etc.). This incoherent sum implies that the sensitivity curve is
obtained by adding together only quadratic terms, averaging to zero the
"interference" terms. In other words, it is assumed that different sources
of noise are not correlated among each other and therefore,%
\begin{equation}
\left\vert \tilde{h}_{\text{sensitivity}}\right\vert ^{2}\overset{\text{def}}%
{=}\sum_{k}\left\langle \tilde{h}_{\text{noise, }k}\tilde{h}_{\text{noise, }%
k}^{\ast }\right\rangle \text{, }
\end{equation}%
where,%
\begin{equation}
\left\langle \tilde{h}_{\text{noise, }i}\tilde{h}_{\text{noise, }j}^{\ast
}\right\rangle =0\text{ if }i\neq j\text{.}
\end{equation}%
Gravity gradients were first identified as a potential noise source in
interferometric gravitational-wave detectors by Weiss \cite{weiss}. The
first quantitative analyses of such gravity-gradient noise were performed by
Saulson \cite{saulson} and Spero \cite{spero}. Improvements of Saulson's
works were carried out by Thorne and others \cite{thorne}. Finally, T.
Creighton at Caltech has revisited the relevance of gravity gradient noise
due to atmospheric fluctuations \cite{creighton}.

In this article, we present a theoretical estimate of the atmospheric
Newtonian noise generated by fluctuations of atmospheric mass densities
generated by acoustic and turbulent phenomena and we determine the relevance
of such noise in the laser-interferometric detection of gravitational waves 
\cite{cafaro}. The structure of this paper is as follows: in Section II, we
briefly review Creighton's work on gravity gradient noise generated by
atmospheric mass fluctuations. In Section III, we present the general
expression for the spectrum of an arbitrary Newtonian noise. In Section IV,
we consider the gravitational coupling of interferometer test-masses to
fluctuations of atmospheric density due to the propagation of sound waves in
a semispace occupied by an ideal fluid delimited by an infinitely rigid
plane. We present an analytical expression of the spectrum of acceleration
fluctuations $\mathcal{S}_{\vec{a}}\left( \vec{r}_{1}\text{, }\vec{r}_{2}%
\text{; }\omega \right) $ of the test-masses of the interferometer in terms
of the experimentally obtainable spectrum of pressure fluctuations $\mathcal{%
S}_{p}\left( \vec{r}_{1}\text{, }\vec{r}_{2}\text{; }\omega \right) $. In
Section V, we consider the gravitational coupling of interferometer
test-masses to fluctuations of atmospheric density due to the propagation of
sound waves generated in a turbulent Lighthill process. We present an
analytical expression - in the Fourier space - of the spectrum of
acceleration fluctuations $\mathcal{\tilde{S}}_{\vec{a}}\left( \vec{k}_{1}%
\text{, }\vec{k}_{2}\text{; }\omega \right) $\ of the test-masses of the
interferometer. Finally, conclusions and final remarks are presented in
Section VI.

\section{Gravity gradient noise due to atmospheric fluctuations: Creighton's
work}

Mass density fluctuations in the atmosphere caused, for example, by acoustic
pressure waves or by\ temperature perturbations induce a stochastic
gravitational field that couples directly to the test masses of
gravitational interferometers and produce noise, the so-called atmospheric
Newtonian noise. The first to discuss the effects of such noise in a
quantitative way was Saulson \cite{saulson}. He considered the effects of
background acoustic pressure waves on the one hand and the motion of massive
bodies in the proximity of the interferometer on the other. Saulson
concluded the atmospheric Newtonian noise he considered would be
insignificant even when using advanced gravitational interferometric
detectors. Later, T. Creighton \cite{creighton} revisited the issue of
atmospheric Newtonian noise. Among the possible sources of mass density
fluctuations in the atmosphere, Creighton considered those that he thought
potentially most significant. The sources he considered were:

\begin{itemize}
\item Low pressure acoustic waves.

\item Fast moving massive bodies in the proximity of the interferometer.

\item Transient atmospheric shock waves.

\item Massive bodies colliding with the ground or the structures housing the
interferometer test masses.

\item Perturbations of atmospheric temperature in the vicinity of the
detector.
\end{itemize}

The largest small scale atmospheric density perturbations are not caused by
pressure waves but rather by temperature perturbations. When heat is
transported through a convective atmospheric layer, convective turbulence
mixes warm and cold air leading to temperature perturbations on all scales
up to the order of millimeters. On the time scales of interest, these
perturbations can be viewed as essentially "frozen"\ effects in the air
mass, while the pressure variations are dispersed quickly via sound waves.
Therefore, the fluctuations in air density are induced primarily from
temperature perturbations which are generally larger than pressure
perturbations by several orders of magnitude. Although "frozen"\ in the air
mass, these temperature perturbations can cause density fluctuations that
vary quickly in time, $d\rho =-\rho \frac{dT}{T}$, when the wind transports
them in space. Indeed, this is the primary source of noise in optical
astronomy. We remark that in this context, it is also important to study the
noise generated by acoustic perturbations (density fluctuations) that are
either completely absorbed or reflected by the ground and/or structures in
the vicinity of the interferometer.

Even though temperature perturbations transported via the wind are the
dominant source of atmospheric density fluctuations, they do not produce
significant Newtonian noise at very high frequencies. This is due to the
long time intervals\textbf{\ }that packets of warm and cold air stay in the
vicinity of the interferometer test masses. A possible exception exist when
the air flow forms vortices around the structure in which the interferometer
is housed. Since such flows can produce a noise spectrum that has a maximum
around the frequencies of the typical vortex circulating in the proximity of
the test mass, the presence of temperature fluctuations in the atmosphere
transported by the wind near the detector can give rise to a non-trivial
source of density fluctuations. Unexpected variations of pressure caused by
atmospheric shock waves can be potential sources of transient signals that
can be detected by the experimental apparatus, provided such phenomena
occurs in the proximity of the detector. These "shocks"\ are relevant
because they can produce significant pressure variations on time scales
smaller than $0.1$ sec. This time scale corresponds to the smallest value of
the band-pass of the majority of current interferometric detectors of
gravitational waves. Such shocks are essentially transient phenomena that
may produce spurious signals in the detector, rather than raise the noise
threshold. It would be useful to know the signal to noise ratio (SNR) that
various shocks could produce.

Although atmospheric shock waves are potential sources of spurious signals
in gravitational wave detectors, they are readily treatable using
environmental detectors. If such sensors record a pressure variation larger
than some millibar on time scales of $50-100$ milliseconds, then we could
expect spurious signals with (adimensional) amplitude of the order of $%
10^{-22}$ in the range of frequencies $10-20$ Hz. An example of atmospheric
shocks is "sonic booms"\ generated by supersonic bodies (for example a
supersonic airplane flying over the space around the detector); they could
obscure the detection of gravitational waves. Even though such events are
rare, if not entirely non-existent, they accentuate the potential
seriousness of shock waves with respect to other weaker acoustic sources or
from acoustic sources originating at greater distances from the experimental
apparatus.

Another potential source of spurious signals in the interferometer is the
Newtonian noise caused by the motion of a single massive body in the
proximity of the detector, or the collision of such a body with the
experimental apparatus. This last possibility is particularly worrisome,
since the deceleration of the body can produce high frequency signals. Next
we discuss one mechanisms that generate atmospheric Newtonian noise from
background acoustic pressure waves.

Consider a pressure plane wave with frequency $f$, propagating at the speed
of sound characteristic of the medium, $c_{s}$. To estimate the relevance of
this noise requires comparison of the spectral density of the acoustic noise
signal to the sensitivity curve of the interferometric detector. The
spectral density of the particular form of noise being considered is given
by \cite{creighton}:%
\begin{equation}
\mathcal{S}_{h}\left( f\right) =\left( \frac{G\rho c}{4\pi ^{2}\gamma l}%
\right) ^{2}\frac{1}{3f^{6}p^{2}}\overset{4}{\underset{i=1}{\sum }}C\left( 
\frac{2\pi fr_{\text{min}}^{\left( i\right) }}{c_{s}}\right) \mathcal{S}%
_{p}^{\left( i\right) }\left( f\right) \text{,}  \label{spect-density}
\end{equation}%
where the indices $i$ identify the various test masses of the
interferometer, $r_{\text{min}}^{\left( i\right) }$ is the dead air radius
about the $i$th test mass, and $\mathcal{S}_{p}^{\left( i\right) }\left(
f\right) $ is the acoustic noise spectrum measured outside the building
enclosing the $i$th test mass. Moreover, $l$ is the length of the
interferometer arm, $\rho $ is the ambient air density, $p$ is the ambient
air pressure and $\gamma =\frac{c_{P}}{c_{V}}$ is the ratio of heat
capacities at constant pressure and temperature of the air at room
temperature. In what follows we briefly reiterate the important points that
led to (\ref{spect-density}).

\begin{enumerate}
\item One supposes that the relative fluctuation of air pressure is small: $%
\frac{\delta p}{p}\ll 1$.

\item Only sound waves in proximity of the interferometer are considered.
Since the interferometer is sensitive only to movement of test-masses
parallel to the arms of the detector, the gravitational acceleration
produced by pressure waves on the test masses is reduced by a factor $\cos
\theta $. Here $\theta $ is the angle between the direction of propagation
of the pressure wave and the arm of the interferometer.

\item The test masses of the interferometer is inside a structure that in
principle can be used to eliminate noise within a characteristic distance $%
r_{min}$ from the test masses. In order to take this into account, a
function $C(x)$ is introduced. The function $C(x)$ depends on the shape of
the structure, the ways in which it reflects sound waves, and on many other
factors.

\item The interferometer is located on the ground, not in an ideal
homogenous empty space. It is assumed that the waves are almost completely
reflected from the ground.
\end{enumerate}

Exploiting these four points, the gravitational acceleration produced by
sound waves in the propagation direction $z$ is:%
\begin{equation}
g_{z}\left( t\right) =G\dint z\frac{\delta \rho \left( t\right) }{r^{3}}dV=%
\frac{G\rho c_{s}}{\gamma pf}\cos \theta \cdot C\left( \frac{2\pi fr_{\text{%
min}}^{\left( i\right) }}{c_{s}}\right) \delta p\left( t+\frac{1}{4f}\right) 
\text{.}  \label{grav-acc}
\end{equation}%
On the other hand, the gravitational wave signal $h\left( t\right) $ in the
interferometer is related to the acceleration of one of the test masses
according to%
\begin{equation}
\frac{d^{2}h}{dt^{2}}=\frac{g\left( t\right) }{l}\text{.}
\end{equation}%
In frequency space this relation becomes%
\begin{equation}
\tilde{h}\left( f\right) =\frac{1}{\left( 2\pi f\right) ^{2}}\frac{\tilde{g}%
\left( f\right) }{l}\text{.}  \label{htilde}
\end{equation}%
Combining (\ref{grav-acc}) with (\ref{htilde}) we obtain%
\begin{equation}
\tilde{h}\left( f\right) =\frac{G\rho c_{s}}{4\pi ^{2}\gamma plf^{3}}\cos
\theta \cdot C\left( \frac{2\pi fr_{\text{min}}^{\left( i\right) }}{c_{s}}%
\right) i\delta \tilde{p}\left( f\right) \text{.}
\end{equation}%
Assuming the noise is stationary and that the directions and amplitudes of
the wave modes are uncorrelated, we have%
\begin{equation}
\mathcal{S}_{h}\left( f\right) =\left[ \frac{G\rho c_{s}}{4\pi ^{2}\gamma
lf^{3}}C\left( \frac{2\pi fr_{\text{min}}^{\left( i\right) }}{c_{s}}\right) %
\right] ^{2}\left\langle \cos ^{2}\theta \right\rangle \frac{\mathcal{S}%
_{p}\left( f\right) }{p^{2}}\text{,}
\end{equation}%
where $\left\langle \cdot \cdot \cdot \right\rangle $ indicates an average
over the modes of the plane wave that contribute to the noise. Finally, we
assume that the test masses in the interferometer are at a distance equal to
several coherent lengths such that the noise can be considered uncorrelated
and therefore, can be summed linearly. Finally using $\left\langle \cos
^{2}\theta \right\rangle =1/3$ we obtain (\ref{spect-density}). Creighton 
\cite{creighton} concludes that the acoustic background and the temperature
fluctuations produce Newtonian noise that is below the sensitivity threshold
of even the most advanced interferometric detectors even though temperature
perturbations transported along streamlines of non-laminar flux could
produce noise within one order of magnitude from the maximum sensitivity
region at $10$ Hz.

A more comprehensive study of such phenomena requires the use of more
sophisticated models. Additionally, shock waves in the atmosphere could
produce potentially significant spurious signals for an advanced
interferometer. These signals could be monitored from acoustic sensors
places outside the interferometer structure. In order to avoid the
non-negligible noise of massive objects transported by the wind and to
preserve the assumption of linear additivity of noise (hypothesis of
uncorrelated noise), it would be necessary to construct barriers to keep
such massive bodies at a secure distance from the test masses. One of the
main purposes of the Virgo project is that of achieving good sensitivity at
low frequencies (around $4-10$ Hz). In this frequency band the thermal and
seismic Newtonian noise represent the dominant sources of noise. If the
thermal noise could be reduced at low frequencies using cryogenic techniques
or by using high Q materials, the seismic Newtonian noise would represent
the sensitivity limit at these frequencies \cite{barone}.

The objective of this article is to present an analytical estimate of the
atmospheric Newtonian noise generated by fluctuations of atmospheric mass
densities and to judge the relevance of this noise in the detection of
gravitational waves using laser-interferometric techniques. With the
development of very refined and sensitive experimental techniques it is
possible to study directly fluctuation phenomena in various areas of
physics. In our view, fluctuation phenomena are related in a natural way to
transport or convection phenomena. Diffusion phenomena and irreversible
thermodynamics are based on results derived from the theory of fluctuations.
We consider non-quantum fluctuations ($\hbar \omega <<k_{B}T$) \cite%
{landau-fluid}, which is the conventional hypothesis employed in fluid
mechanics.

Additionally, we suppose that the viscosity coefficients and thermal
conductivity of the fluid (atmosphere) are non-dispersive, that is, they are
independent of the oscillations of fluctuation $\omega $. The phenomena that
cause such fluctuations of atmospheric mass density are divided into two
groups:

\begin{enumerate}
\item Fluctuations generated from Acoustic Phenomena.

\item Fluctuations generated from Turbulent Phenomena.
\end{enumerate}

We study theoretical models corresponding to each of these groups and
present estimates of the spectral density of the atmospheric Newtonian noise
for each case.

\section{Calculation of Newtonian Noise}

Before studying these physical phenomena, we first calculate some relevant
quantities that will enable us to quantitatively describe the shape of the
Newtonian noise. The effect of atmospheric density fluctuations $\delta \rho
\left( \vec{x}\text{, }t\right) $ on each one of the four test-masses that
are suspended on the suspension tower of the interferometer, can be evaluate
using Newton's force law $\vec{F}=m\vec{a}$ with%
\begin{equation}
\vec{a}\left( \vec{x}\text{, }t\right) =-G\int d^{3}\vec{x}^{\prime }\delta
\rho \left( \vec{x}^{\prime }\text{, }t\right) \frac{\vec{x}-\vec{x}^{\prime
}}{\left\Vert \vec{x}-\vec{x}^{\prime }\right\Vert ^{3}}
\end{equation}%
where $\vec{x}$ is the effective position of the test-mass. The test-masses
are arranged so as to form the two arms of the interferometer where these
arms are oriented along the $x$ and $y$ directions. Furthermore, since only
variations of the relative length of the light path are detected, we are
only interested in the $x$ and $y$ components of the acceleration. The
relative difference of the arm length is given by $h\left( t\right) $ 
\begin{equation}
h\left( t\right) =\frac{1}{L}\left[ \left( x_{2}^{\left( 1\right)
}-x_{2}^{\left( 2\right) }\right) -\left( x_{1}^{\left( 4\right)
}-x_{1}^{\left( 3\right) }\right) \right] \text{,}
\end{equation}%
where $L$ is the length of the arms of the interferometer. Therefore, we can
link the temporal second derivative of $h\left( t\right) $ to the Newtonian
acceleration of the masses test, 
\begin{equation}
\frac{d^{2}h\left( t\right) }{dt^{2}}=\frac{1}{L}\left[ \left( a_{2}^{\left(
1\right) }-a_{2}^{\left( 2\right) }\right) -\left( a_{1}^{\left( 4\right)
}-a_{1}^{\left( 3\right) }\right) \right] \text{.}  \label{h2}
\end{equation}%
Since we are interested in the spectral amplitude of the atmospheric
Newtonian noise, let us consider the Fourier transform of $h\left( t\right) $%
, expressed formally as 
\begin{equation}
\tilde{h}\left( \omega \right) =\int dth\left( t\right) e^{i\omega t}\text{.}
\end{equation}%
In the Fourier space, (\ref{h2}) becomes%
\begin{equation}
-\omega ^{2}\tilde{h}\left( \omega \right) =\frac{1}{L}\left[ \left( \tilde{a%
}_{2}^{\left( 1\right) }-\tilde{a}_{2}^{\left( 2\right) }\right) -\left( 
\tilde{a}_{1}^{\left( 4\right) }-\tilde{a}_{1}^{\left( 3\right) }\right) %
\right] \text{.}  \label{you}
\end{equation}%
Due to the stochastic nature of the density fluctuations, we are concerned
with the ensemble average $\mathcal{S}_{h}\left( \omega \right) $ defined as%
\begin{equation}
\mathcal{S}_{h}\left( \omega \right) =\left\langle \tilde{h}\left( \omega
\right) \tilde{h}^{\ast }\left( \omega \right) \right\rangle \text{.}
\label{ens-avg}
\end{equation}%
Equation (\ref{ens-avg}) represents the spectrum of Newtonian noise. Using (%
\ref{you}), (\ref{ens-avg}) becomes%
\begin{equation}
\mathcal{S}_{h}\left( \omega \right) =\frac{1}{L^{2}\omega ^{4}}\left\langle
\left\vert \left( \tilde{a}_{2}^{\left( 1\right) }-\tilde{a}_{2}^{\left(
2\right) }\right) -\left( \tilde{a}_{1}^{\left( 4\right) }-\tilde{a}%
_{1}^{\left( 3\right) }\right) \right\vert ^{2}\right\rangle \text{.}
\label{ens-avg2}
\end{equation}%
We point out that (\ref{ens-avg2}) will be simplified in a convenient manner
in the cases we consider by assuming conditions of homogeneity (invariance
under translations in a statistical sense) and isotropy (invariance under
rotation in a statistical sense) of the correlation functions describing
density fluctuations and therefore, the fluctuations of acceleration.

\section{Acoustic Phenomena}

It is the interplay between the compressibility and inertia of the fluid
that supports the propagation of sound waves (the oscillating motion of
small amplitudes in an incompressible fluid is called sound waves) in the
medium. We work with the linear theory of acoustics since we consider
perturbations that are negligible in the equations of motion. We consider
exclusively the compressibility and inertia of the fluid, but no other
property of the fluid. We obtain the linearized equations of the acoustic
theory in their simplest non-trivial form. We neglect the influence on the
propagation of sound waves from viscosity, heat conduction and inhomogeneity
at the boundaries. We consider the propagation of waves in an ideal fluid,
with reference to so-called "small motions". That is to say, we consider a
linearized approximation of the Euler equation \cite{landau-fluid}%
\begin{equation}
\frac{\partial \vec{v}}{\partial t}+(\vec{v}\cdot \vec{\nabla}\vec{v})=-%
\frac{1}{\rho }\vec{\nabla}p\text{.}  \label{euler}
\end{equation}%
This procedure is legitimate when the perturbations in pressure and density (%
$\delta p$ and $\delta \rho $, respectively) are sufficiently small compared
to the equilibrium state ($p_{0}$ and $\rho _{0}$, respectively). We proceed
to linearize (\ref{euler}) together with the continuity equation%
\begin{equation}
\frac{\partial \rho }{\partial t}+\vec{\nabla}\cdot \left( \rho \vec{v}%
\right) =0  \label{11}
\end{equation}%
and the adiabatic condition 
\begin{equation}
\frac{ds}{dt}=0
\end{equation}%
($s$ it is the entropy for unit mass). Letting $p=p_{0}+\delta p$, $\rho
=\rho _{0}+\delta \rho $ where $\rho _{0}$ and $p_{0}$ are the density and
pressure of the fluid at equilibrium respectively, $\delta \rho $ and $%
\delta p$ are the perturbations of density and pressure satisfying $\delta
\rho <<\rho _{0}$ and $\delta p<<p_{0}$. Defining the velocity potential
function $\phi $ such that $\vec{\nabla}\phi =\vec{v}$ (in the reasonable
hypothesis that the velocity field is irrotational, $\vec{\nabla}\times \vec{%
v}=0$), we obtain the sound wave equation%
\begin{equation}
\left( \Delta -\frac{1}{c_{s}^{2}}\frac{\partial ^{2}}{\partial t^{2}}%
\right) \phi \left( \vec{x}\text{, }t\right) =0  \label{sound-wave}
\end{equation}%
where $c_{s}=\sqrt{\left( \partial p/\partial \rho \right) _{s}}$ is the
speed of sound and can be expressed as $c_{s}=1/\sqrt{\rho \chi _{s}}$ where 
$\chi _{s}=\rho ^{-1}\left( \partial p/\partial \rho \right) _{s}$ is the
adiabatic compressibility of the medium. It is straightforward to verify
that $\delta \rho $ and $\delta p$ satisfy the same wave equation once we
recognize the relationship between the fluctuation of pressure and density
of a compressible medium is given by $\delta p=c_{s}^{2}\delta \rho $. We
emphasize that the wave equation introduced for $\phi $, $\delta r$ or $%
\delta p$ are linear approximations. We will not consider non-linear wave
propagation phenomena. We recall however, that non-linearities may produce
"distortion"\ of the initial signal or even the emergence of discontinuities
(shock waves). At this juncture we highlight the two fundamental assumptions
that enable us to obtain the linearized equations:

\begin{enumerate}
\item The necessary condition for applying the linearized equations of
motion to describe the propagation of sound waves is that the speed of the
fluid particles is small compared to the speed of sound, $v<<c_{s}$.

\item The propagation of sound waves in an ideal fluid is considered
adiabatic.
\end{enumerate}

It is reasonable to assume that the process of compression and rarefaction
occurring locally is adiabatic even though as a consequence of this thermal
inhomogeneities arise in the medium. Such inhomogeneities would lead to
hypothesize the existence of heat exchange between adjacent regions at
different temperature. It turns out however, that the time scale over which
appreciable heat exchange occurs (locally) is much longer than the time
scale over which the causes (local pressure and temperature gradients) that
generate the thermal exchange are sustained. In other words, we assume%
\begin{equation}
T_{\text{oscillation}}\ll \tau _{\text{conduction}}
\end{equation}%
where $T_{\text{oscillation}}$ represents the period of the sound wave,
while $\tau _{\text{conduction}}$ represents the characteristic time scale
over which a significant heat exchange in the medium can occur. These two
time scales can be numerically estimated, yielding%
\begin{equation}
\tau _{\text{conduction}}\simeq \frac{(\frac{\lambda }{2})^{2}}{\chi }=\frac{%
c_{s}^{2}}{\chi f^{2}}\simeq 10^{10}\left( \frac{1\text{Hz}}{f}\right) ^{2}%
\text{sec.}
\end{equation}%
where $\lambda $ is the wavelength considered and $\chi $ is the
thermometric conductivity of the air. Clearly the conduction time $\tau _{%
\text{conduction}}$ is much greater than any oscillation time $T_{\text{%
oscillation}}$ of interest to us.

\subsection{Propagation of Acoustic Waves in a Semispace}

As a first problem concerning the generation of atmospheric Newtonian noise,
let us consider the gravitational coupling of the interferometer test-masses
to fluctuations of atmospheric density due to the propagation of sound waves
in a semispace occupied by a compressible ideal fluid (atmosphere), where
the semispace is delimited by an infinitely rigid plane (ground). It is
necessary to integrate (\ref{sound-wave}) with the boundary condition 
\begin{equation}
\frac{\partial \phi \left( x\text{, }y\text{, }z=0\text{, }t\right) }{%
\partial z}=0\text{.}
\end{equation}%
This condition expresses the fact that the velocity of the ideal fluid on
the surface of separation is purely tangential. The general solution of this
problem is given by%
\begin{equation}
\phi \left( \vec{x}\text{, }t\right) =\int d^{3}\vec{k}f(\vec{k})u_{\vec{k}%
}\left( \vec{x}\text{, }t\right)
\end{equation}%
where $u_{\vec{k}}$ are plane waves defined as%
\begin{equation}
u_{\vec{k}}\left( \vec{x}\text{, }t\right) =\exp \left[ i(\vec{k}\cdot \vec{x%
}-\omega t)\right] \text{.}
\end{equation}%
The function $f(\vec{k})$ is defined such that it satisfies $f\left( k_{x}%
\text{, }k_{y}\text{, }k_{z}\right) =f\left( k_{x}\text{, }k_{y}\text{, }%
-k_{z}\right) $ (to ensure normalization) as well as the boundary conditions
of the problem. In order to find expressions for the acoustic Newtonian
noise it is necessary to calculate the correlation function associated with
the velocity potential $\phi $ relative to fluctuations of acceleration of
the test-masses. From fluid mechanics \cite{landau-fluid} we find%
\begin{equation}
\delta \rho \left( \vec{x}\text{, }t\right) =-\frac{\rho _{0}}{c_{s}^{2}}%
\frac{\partial \phi \left( \vec{x}\text{, }t\right) }{\partial t}\text{.}
\label{inter1}
\end{equation}%
In Fourier space (\ref{inter1}) becomes%
\begin{equation}
\delta \tilde{\rho}\left( \vec{x}\text{, }\omega \right) =\frac{i\omega \rho
_{0}}{c_{s}^{2}}\tilde{\phi}\left( \vec{x}\text{, }\omega \right) \text{.}
\label{inter2}
\end{equation}%
Multiplying (\ref{inter2}) for its complex conjugate and averaging (ensemble
average) we obtain,%
\begin{equation}
\mathcal{S}_{\delta \rho }\left( \vec{x}_{1}\text{, }\vec{x}_{2}\text{; }%
\omega \right) =\left\langle \delta \tilde{\rho}\left( \vec{x}_{1}\text{, }%
\omega \right) \delta \tilde{\rho}^{\ast }\left( \vec{x}_{2}\text{, }\omega
\right) \right\rangle =\frac{\omega ^{2}\rho _{0}^{2}}{c_{s}^{2}}%
\left\langle \tilde{\phi}\left( \vec{x}_{1}\text{, }\omega \right) \tilde{%
\phi}^{\ast }\left( \vec{x}_{2}\text{, }\omega \right) \right\rangle \text{.}
\label{inter3}
\end{equation}%
Furthermore, since%
\begin{equation}
a_{i}\left( \vec{x}\text{, }t\right) =-G\int d^{3}\vec{x}^{\prime }\delta
\rho \left( \vec{x}^{\prime }\text{, }t\right) \frac{x_{i}-x_{i}^{\prime }}{%
\left\Vert \vec{x}-\vec{x}^{\prime }\right\Vert ^{3}}\text{,}
\end{equation}%
we find, proceeding as in (\ref{inter2}),%
\begin{eqnarray}
\left( \mathcal{S}_{a}\right) _{ij}\left( \vec{x}_{1}\text{, }\vec{x}_{2}%
\text{; }\omega \right) &=&\left\langle \tilde{a}_{i}\left( \vec{x}_{1}\text{%
, }\omega \right) \tilde{a}_{j}^{\ast }\left( \vec{x}_{2}\text{, }\omega
\right) \right\rangle  \label{inter4} \\
&&  \notag \\
&=&G\int d^{3}\vec{x}_{1}^{\prime }\int d^{3}\vec{x}_{2}^{\prime }\frac{%
\left( \vec{x}_{1}-\vec{x}_{1}^{\prime }\right) _{i}\left( \vec{x}_{2}-\vec{x%
}_{2}^{\prime }\right) _{j}}{\left\Vert \vec{x}_{1}^{\prime }-\vec{x}%
_{1}\right\Vert ^{3}\left\Vert \vec{x}_{2}^{\prime }-\vec{x}_{2}\right\Vert
^{3}}\mathcal{S}_{\delta \rho }\left( \vec{x}_{1}^{\prime }\text{, }\vec{x}%
_{2}^{\prime }\text{; }\omega \right)  \notag
\end{eqnarray}%
where $a_{i}\left( \vec{x}\text{, }t\right) $ is the $i^{\text{th}}$
component of the fluctuation of acceleration exerted on the test-mass
located at $\vec{x}$. This fluctuation of acceleration is generated by
fluctuations of atmospheric mass density occupying volume $V$. For economy
of notation we define the quantity%
\begin{equation}
\mathcal{K}_{ij}\left( \vec{x}\text{, }\vec{y}\right) =\frac{x_{i}y_{j}}{%
\left\Vert \vec{x}\right\Vert ^{3}\left\Vert \vec{y}\right\Vert ^{3}}\text{.}
\label{yy}
\end{equation}%
Replacing (\ref{inter3}) and (\ref{yy}) in (\ref{inter4}), we obtain%
\begin{equation}
\left\langle \tilde{a}_{i}\left( \vec{x}_{1}\text{, }\omega \right) \tilde{a}%
_{j}^{\ast }\left( \vec{x}_{2}\text{, }\omega \right) \right\rangle =\frac{%
G^{2}\omega ^{2}\rho _{0}^{2}}{c_{s}^{4}}\int d^{3}\vec{x}_{1}^{\prime }\int
d^{3}\vec{x}_{2}^{\prime }\mathcal{K}_{ij}\left( \vec{x}_{1}^{\prime }-\vec{x%
}_{1}\text{, }\vec{x}_{2}^{\prime }-\vec{x}_{2}\right) \left\langle \tilde{%
\phi}\left( \vec{x}_{1}^{\prime }\text{, }\omega \right) \tilde{\phi}^{\ast
}\left( \vec{x}_{2}^{\prime }\text{, }\omega \right) \right\rangle \text{.}
\end{equation}%
Therefore the main problem is that of evaluating the expression for $S_{\phi
}\left( \vec{x}_{1}\text{, }\vec{x}_{2}\text{, }\omega \right) $,%
\begin{equation}
\mathcal{S}_{\phi }\left( \vec{x}_{1}\text{, }\vec{x}_{2}\text{; }\omega
\right) =\left\langle \tilde{\phi}\left( \vec{x}_{1}\text{, }\omega \right) 
\tilde{\phi}^{\ast }\left( \vec{x}_{2}\text{, }\omega \right) \right\rangle
=\int d^{3}\vec{k}_{1}\int d^{3}\vec{k}_{2}\tilde{u}_{\vec{k}_{1}}\left( 
\vec{x}_{1}\text{, }\omega \right) \tilde{u}_{\vec{k}_{2}}^{\ast }\left( 
\vec{x}_{2}\text{, }\omega \right) \left\langle f(\vec{k}_{1})f^{\ast }(\vec{%
k}_{2})\right\rangle  \label{inter6}
\end{equation}%
or in other words, that of evaluating the correlation function $\left\langle
f\left( \vec{k}_{1}\right) f^{\ast }\left( \vec{k}_{2}\right) \right\rangle $%
. Assuming this correlation function is homogenous and isotropic, that is to
say, assuming invariance (in a statistical sense) under translations and
rotations, we have%
\begin{equation}
\left\langle f(\vec{k}_{1})f^{\ast }(\vec{k}_{2})\right\rangle =(2\pi
)^{3}\delta ^{\left( 3\right) }\left( \vec{k}_{1}-\vec{k}_{2}\right)
\left\langle \left\vert f(\vec{k}_{1})\right\vert ^{2}\right\rangle \text{.}
\label{inter5}
\end{equation}%
Including the constants of normalization in the definition of $f(\vec{k})$
and substituting (\ref{inter5}) into (\ref{inter6}) we arrive at%
\begin{equation}
\left\langle \tilde{\phi}\left( \vec{x}_{1}\text{, }\omega \right) \tilde{%
\phi}^{\ast }\left( \vec{x}_{2}\text{, }\omega \right) \right\rangle =\int
d^{3}\vec{k}\tilde{u}_{\vec{k}}\left( \vec{x}_{1}\text{, }\omega \right) 
\tilde{u}_{\vec{k}}^{\ast }\left( \vec{x}_{2}\text{, }\omega \right)
\left\langle \left\vert f(\vec{k})\right\vert ^{2}\right\rangle \text{.}
\end{equation}%
The problem has now been reduced to the calculation of the quantity$%
\left\langle \left\vert f\left( \vec{k}\right) \right\vert ^{2}\right\rangle 
$. From the general expression for $f$ and using $\vec{v}=\vec{\nabla}\phi $%
, we get%
\begin{equation}
\vec{\nabla}\cdot \vec{v}=-\int d^{3}\vec{k}k^{2}f(\vec{k})u_{\vec{k}}\left( 
\vec{x}\text{, }t\right) \text{.}  \label{1}
\end{equation}%
The continuity equation (\ref{11}) expressed in frequency space reads%
\begin{equation}
\delta \tilde{\rho}\left( \vec{x}\text{, }\omega \right) =\frac{-i\rho _{0}}{%
\omega }\vec{\nabla}\cdot \vec{v}\text{.}  \label{inter7}
\end{equation}%
Assuming fixed $\omega $ and using (\ref{1}), equation (\ref{inter7}) becomes%
\begin{equation}
\delta \tilde{\rho}\left( \vec{x}\text{, }\omega \right) =\frac{i\rho
_{0}\omega ^{3}}{c_{s}^{4}}\int d\Omega _{\vec{k}}f(\vec{k})u_{\vec{k}%
}\left( \vec{x}\right)
\end{equation}%
and therefore%
\begin{equation}
\left\langle \delta \tilde{\rho}\left( \vec{x}_{1}\text{, }\omega \right)
\delta \tilde{\rho}^{\ast }\left( \vec{x}_{2}\text{, }\omega \right)
\right\rangle =\frac{\rho _{0}^{2}\omega ^{6}}{c_{s}^{8}}\int d\Omega _{\vec{%
k}}\int d\Omega _{\vec{p}}u_{\vec{k}}\left( \vec{x}_{1}\right) u_{\vec{p}%
}^{\ast }\left( \vec{x}_{2}\right) \left\langle f(\vec{k})f^{\ast }(\vec{p}%
)\right\rangle \text{.}  \label{inter8}
\end{equation}%
Using the invariance under rotation around the $z$ axis and the constraint
conditions on the amplitudes $f(\vec{k})$, it is found that (\ref{inter8})
becomes%
\begin{equation}
\mathcal{S}_{\delta \rho }\left( \vec{x}_{1}\text{, }\vec{x}_{2}\text{; }%
\omega \right) =\frac{2\rho _{0}^{2}\omega ^{6}}{c_{s}^{8}}%
\dint\limits_{0}^{1}d\xi \mathcal{F}\left( \xi \text{, }\omega \right)
J_{0}\left( \frac{\omega r}{c_{s}}\sqrt{1-\xi ^{2}}\right) \left\{ \cos %
\left[ \frac{\omega \xi }{c_{s}}\left( z_{1}-z_{2}\right) \right] +\cos %
\left[ \frac{\omega \xi }{c_{s}}\left( z_{1}+z_{2}\right) \right] \right\} 
\text{,}  \label{inter9}
\end{equation}%
where $r=\sqrt{\left( x_{1}-x_{2}\right) ^{2}+\left( y_{1}-y_{2}\right) ^{2}}
$, $\mathcal{F}\left( \xi \text{, }\omega \right) =\left\langle \left\vert
f\left( \xi \text{, }\omega \right) \right\vert ^{2}\right\rangle $ and $%
J_{0}\left( z\right) $ is the Bessel function of the first kind \cite{morse}%
. Assuming $\vec{x}_{1}=\vec{x}_{2}=\left( 0\text{, }0\text{, }z\right) $,
equation (\ref{inter9}) simplifies and becomes%
\begin{equation}
\mathcal{S}_{\delta \rho }\left( z\text{, }\omega \right) =\frac{2\rho
_{0}^{2}\omega ^{6}}{c_{s}^{8}}\dint\limits_{0}^{1}d\xi \mathcal{F}\left(
\xi \text{, }\omega \right) \left[ 1+\cos \left( \frac{2\omega \xi z}{c_{s}}%
\right) \right] \text{.}
\end{equation}%
Since $c_{s}^{4}\mathcal{S}_{\delta \rho }=\mathcal{S}_{p}$, we obtain%
\begin{equation}
\mathcal{S}_{p}\left( z\text{, }\omega \right) =\frac{2\rho _{0}^{2}\omega
^{6}}{c_{s}^{4}}\dint\limits_{0}^{1}d\xi \mathcal{F}\left( \xi \text{, }%
\omega \right) \left[ 1+\cos \left( \frac{2\omega \xi z}{c_{s}}\right) %
\right] \text{.}  \label{inter10}
\end{equation}%
From (\ref{inter10}) we get%
\begin{equation}
\dint\limits_{0}^{+\infty }dz\cos \left( \alpha z\right) \mathcal{S}%
_{p}\left( z\text{, }\omega \right) =\frac{2\rho _{0}^{2}\omega ^{6}}{%
c_{s}^{4}}\dint\limits_{0}^{1}d\xi \mathcal{F}\left( \xi \text{, }\omega
\right) \dint\limits_{0}^{+\infty }dz\cos \alpha z\left[ 1+\cos \left( \frac{%
2\omega \xi z}{c_{s}}\right) \right] \text{,}
\end{equation}%
that is%
\begin{equation}
\dint\limits_{0}^{+\infty }dz\cos \left( \alpha z\right) \mathcal{S}%
_{p}\left( z\text{, }\omega \right) =\frac{\pi \rho _{0}^{2}\omega ^{6}}{%
c_{s}^{4}}\mathcal{F}\left( \frac{c_{s}}{2\omega }\alpha \text{, }\omega
\right) \text{.}
\end{equation}%
We therefore conclude,%
\begin{equation}
\mathcal{F}\left( \frac{c_{s}}{2\omega }\alpha \text{, }\omega \right) =%
\frac{c_{s}^{4}}{\pi \rho _{0}^{2}\omega ^{6}}\dint\limits_{0}^{+\infty
}dz\cos \left( \alpha z\right) \mathcal{S}_{p}\left( z\text{, }\omega
\right) \text{.}  \label{mmk}
\end{equation}%
The spectrum of pressure $\mathcal{S}_{p}(z$, $\omega )$ can be obtained
experimentally by use of an acoustic detector (microphone) directed along
different directions (the parameter $\alpha $ accounts for this directional
variability of the acoustic detector). Hence, by integrating (\ref{mmk}) in $%
z$, the quantity $\mathcal{F}\left( \frac{c_{s}}{2\omega }\alpha \text{, }%
\omega \right) $ can be readily calculated. Using $\mathcal{F}\left( \frac{%
c_{s}}{2\omega }\alpha \text{, }\omega \right) $ it is possible to estimate
the correlation function associated with fluctuations of acceleration of the
test-masses. Alternatively, it is possible to determine any correlation
starting from the correlations at the ground level. A simple way to verify
this consists of applying a Fourier transform to the variables $x$, $y$, $t$
in the wave equations and boundary conditions. The problem is then reduced
to integration of an ordinary differential equation with initial conditions
depending on the values of pressure at the ground level. Due to a lack of
detailed measurements of the correlations of pressure, it is necessary to
further simplify the model. For example, we can start again from the wave
equation for pressure fluctuation and Fourier transform only the time
variable,%
\begin{equation}
\left\{ 
\begin{array}{c}
\left( \nabla ^{2}+\frac{\omega ^{2}}{c_{s}^{2}}\right) \delta \tilde{p}%
\left( \vec{r}\text{, }\omega \right) =0 \\ 
\\ 
\left( \frac{\partial \delta \tilde{p}\left( \vec{r}\text{, }\omega \right) 
}{\partial z}\right) _{z=0}=0 \\ 
\\ 
\left[ \delta \tilde{p}\left( \vec{r}\text{, }\omega \right) \right]
_{z=0}=\delta \tilde{p}_{\text{exp.}}\left( x\text{, }y\text{, }\omega
\right)%
\end{array}%
\right.  \label{inter11}
\end{equation}%
In the boundary conditions appears measures of the fluctuations at ground
level, $\delta \tilde{p}_{\text{exp}}$. The solution of problem (\ref%
{inter11}) can be written as the sum on modes at fixed frequency $\omega $,%
\begin{equation}
\delta \tilde{p}\left( \vec{r}\text{, }\omega \right) =\int \frac{d^{2}\vec{k%
}}{\left( 2\pi \right) ^{2}}\mathcal{A}\left( \vec{k}\text{, }\omega \right)
e^{i\vec{k}\cdot \vec{r}}\cos \left( \gamma _{\vec{k}}z\right)
\end{equation}%
where $\gamma _{\vec{k}}=\sqrt{\omega ^{2}/c_{s}^{2}-k^{2}}$ and the
integration is extended to all values of $\vec{k}=\left( k_{x}\text{, }k_{y}%
\text{, }0\right) $ such that $\gamma _{\vec{k}}$ is real. Then the
correlations of pressure fluctuations can be written as%
\begin{equation}
\mathcal{S}_{p}\left( \vec{r}_{1}\text{, }\vec{r}_{2}\text{; }\omega \right)
=\int \frac{d^{2}\vec{k}_{1}}{\left( 2\pi \right) ^{2}}\int \frac{d^{2}\vec{k%
}_{2}}{\left( 2\pi \right) ^{2}}\left\langle \mathcal{A}\left( \vec{k}_{1}%
\text{, }\omega \right) \mathcal{A}^{\ast }\left( \vec{k}_{2}\text{, }\omega
\right) \right\rangle e^{i\vec{k}_{1}\cdot \vec{r}_{1}}e^{-i\vec{k}_{2}\cdot 
\vec{r}_{2}}\cos \left( \gamma _{\vec{k}_{1}}z_{1}\right) \cos \left( \gamma
_{\vec{k}_{2}}z_{2}\right)
\end{equation}%
The simplifying hypothesis is that the correlations between the amplitudes
of modes have the following form%
\begin{equation}
\left\langle \mathcal{A}\left( \vec{k}_{1}\text{, }\omega \right) \mathcal{A}%
^{\ast }\left( \vec{k}_{2}\text{, }\omega \right) \right\rangle =\left( 2\pi
\right) ^{2}\Lambda \left( \omega \right) \delta ^{\left( 2\right) }\left( 
\vec{k}_{1}-\vec{k}_{2}\right)
\end{equation}%
that is to say, different modes are completely uncorrelated and depend only
on the frequency $\omega $. From such hypothesis we can rewrite the
correlations of pressure fluctuations as%
\begin{equation}
\mathcal{S}_{p}\left( \vec{r}_{1}\text{, }\vec{r}_{2}\text{; }\omega \right)
=\frac{1}{2\pi }\left( \frac{\omega }{c_{s}}\right) ^{2}\Lambda \left(
\omega \right) \dint\limits_{0}^{1}d\eta \eta J_{0}\left( \frac{\omega \eta 
}{c_{s}}R_{12}\right) \cos \left( \frac{\omega z_{1}}{c_{s}}\sqrt{1-\eta ^{2}%
}\right) \cos \left( \frac{\omega z_{2}}{c_{s}}\sqrt{1-\eta ^{2}}\right)
\label{inter12}
\end{equation}%
where $R_{12}=\sqrt{\left( x_{1}-x_{2}\right) ^{2}+\left( y_{1}-y_{2}\right)
^{2}}$. At this point we consider a local measurement of the spectrum of
pressure fluctuations given by,%
\begin{equation}
\mathcal{S}_{p}\left( 0\text{, }0\text{, }\omega \right) =\frac{1}{2\pi }%
\left( \frac{\omega }{c_{s}}\right) ^{2}\Lambda \left( \omega \right) \text{.%
}  \label{inter13}
\end{equation}%
Using (\ref{inter12}) and (\ref{inter13}) we obtain the result%
\begin{equation}
\frac{\mathcal{S}_{p}\left( \vec{r}_{1}\text{, }\vec{r}_{2}\text{; }\omega
\right) }{\mathcal{S}_{p}\left( 0\text{, }0\text{; }\omega \right) }%
=\dint\limits_{0}^{1}d\eta \eta J_{0}\left( \frac{\omega \eta }{c_{s}}%
R_{12}\right) \cos \left( \frac{\omega z_{1}}{c_{s}}\sqrt{1-\eta ^{2}}%
\right) \cos \left( \frac{\omega z_{2}}{c_{s}}\sqrt{1-\eta ^{2}}\right) 
\text{.}
\end{equation}%
Finally, for the correlation between accelerations we find%
\begin{eqnarray}
\left( \mathcal{S}_{a}\right) _{ij}\left( \vec{r}_{1}\text{, }\vec{r}_{2}%
\text{; }\omega \right) &=&\frac{G^{2}}{c_{s}^{4}}\mathcal{S}_{p}\left( 0%
\text{, }0\text{; }\omega \right) \int d^{3}\vec{r}_{1}^{\prime }\int d^{3}%
\vec{r}_{2}^{\prime }\mathcal{K}_{ij}\left( \vec{r}_{1}-\vec{r}_{1}^{\prime }%
\text{, }\vec{r}_{2}-\vec{r}_{2}^{\prime }\right) \times  \notag \\
&&  \label{inter14} \\
&&\times \dint\limits_{0}^{1}d\eta \eta J_{0}\left( \frac{\omega \eta }{c_{s}%
}R_{12}^{\prime }\right) \cos \left( \frac{\omega z_{1}^{\prime }}{c_{s}}%
\sqrt{1-\eta ^{2}}\right) \cos \left( \frac{\omega z_{2}^{\prime }}{c_{s}}%
\sqrt{1-\eta ^{2}}\right) \text{,}  \notag
\end{eqnarray}%
where $J_{m}\left( z\right) $ is the Bessel function of the first kind given
by,%
\begin{equation}
J_{m}\left( z\right) \overset{\text{def}}{=}\sum_{k=0}^{\infty }\frac{\left(
-1\right) ^{k}}{2^{2k+\left\vert m\right\vert }k!\left( \left\vert
m\right\vert +k\right) !}z^{2k+\left\vert m\right\vert }\text{, }\left\vert
m\right\vert \neq \frac{1}{2}\text{.}
\end{equation}%
For $m=0$, $J_{0}\left( z\right) $ becomes\textbf{\ }\cite{morse},%
\begin{equation}
J_{0}\left( z\right) =\frac{1}{\pi }\int_{0}^{\pi }\exp \left( iz\cos \theta
\right) d\theta =\sum_{k=0}^{\infty }\left( -1\right) ^{k}\text{\ \ }\frac{%
\left( \frac{1}{4}z^{2}\right) ^{k}}{\left( k!\right) ^{2}}\text{.}
\label{15}
\end{equation}%
Equation (\ref{inter14}) allows to obtain an analytical estimate for the
correlation between the Newtonian accelerations for any pair of points in
terms of the spectrum of pressure fluctuations which can be experimentally
determined. Unfortunately, we do not have direct measurements of $\mathcal{S}%
_{p}$ for the VIRGO detector. As evident from (\ref{inter14}) and (\ref{15})%
\textbf{,} $\left( \mathcal{S}_{a}\right) _{ij}\left( \vec{r}_{1}\text{, }%
\vec{r}_{2}\text{; }\omega \right) $ has a non-trivial dependence on $\omega 
$ and therefore to compare the effect of this noise with the sensitivity
curve of Virgo, it is crucial to focus on the proper frequency-band of the
noise considered. Then, we compare the square root of the strain amplitude $%
\tilde{h}_{rss}\left( f\right) $ of Virgo with the square root of the strain
amplitude of the noise considered. In principle, we could use our analytical
estimate together with experimental values extracted from the literature 
\cite{pp} and provide numerical evidence leading to a numerical estimate of
the relevance of the noise considered.

\section{Turbulent Phenomena}

Before considering our specific problem, it is useful to briefly discuss the
main characteristics of turbulence. Consider a turbulent flow of a
incompressible fluid. A turbulent flow is by definition unstable: a small
perturbation will in general be amplified due to non-linearities appearing
in the equations describing the flow. Furthermore, it is evident from a
great amount of experimental data the turbulent flow of an incompressible
fluid is rotational, that is to say, $\vec{\omega}=\vec{\nabla}\times \vec{v}
$ $\neq 0$, at least in certain regions of the space. The set of equations
that defines this physical system consists is the Navier-Stokes \cite%
{landau-fluid} equation 
\begin{equation}
\frac{\partial \vec{v}}{\partial t}+\left( \vec{v}\cdot \vec{\nabla}\right) 
\vec{v}=-\frac{1}{\rho }\vec{\nabla}p+\nu \vec{\nabla}^{2}\vec{v}
\end{equation}%
where $\nu $ it is the kinematic viscosity of the fluid and the
incompressible condition reads%
\begin{equation}
\vec{\nabla}\cdot \vec{v}=0\text{.}
\end{equation}%
The system should be integrated taking account of the chosen initial and
boundary conditions. The Navier-Stokes equation encodes all we need to know
about turbulence. However, it is essential to have data from experimental
observations in order to properly understand the phenomena since it is a
highly non-trivial task to analytically integrate the equation due to its
inherent non-linearities. Among the parameters that characterize the
turbulent flow, only the kinematic viscosity $\nu =\frac{\eta }{\rho }$ ($%
\eta $ is the dynamic viscosity) appears in the Navier-Stokes equations. The
unknown quantities to be determined are $\vec{v}$ and $\frac{p}{\rho }$.
Moreover the flow depends - through the boundary conditions - on the shape
and dimensions of the body being inserted into the fluid in order to break
the irrotationality of the velocity field thereby giving rise to turbulence.
Since generally the shape of the body is assumed given, the geometric
properties are characterized by a typical linear dimension denoted $l$. Let $%
u$ be a typical speed of the principal flow of the fluid. Then every flow is
specified by three parameters: $\nu $, $u$ and $l$. The only adimensional
quantity that can be constructed from these three parameters is the
so-called Reynolds number $\mathcal{R}$ \cite{landau-fluid},%
\begin{equation}
\mathcal{R}=\frac{ul}{v}\simeq \frac{\left\Vert \left( \vec{v}\cdot \vec{%
\nabla}\right) \vec{v}\right\Vert }{\left\Vert \nu \vec{\nabla}^{2}\vec{v}%
\right\Vert }\text{.}  \label{reynolds}
\end{equation}%
The numerator of (\ref{reynolds}) represents the transport term (or inertial
term), while the denominator is the viscosity term. When the Reynolds number
is small ($\mathcal{R}\ll 1$), it is permissible to neglect the inertial
forces and therefore, the Navier-Stokes equations can be linearized. On the
other hand, when $\mathcal{R}>>1$ (large Reynolds numbers), the inertial
forces dominate those of viscosity. In such situations, instabilities
develop that lead to chaotic motion (turbulence).

In order to clarify the mechanism that leads to the emergence of turbulence,
we introduce the concepts of broken and restored symmetries. The concept of
symmetry is central in the study of transition phenomena and fully developed
turbulence. Transformation symmetries are represented by either continuous
or discreet invariance groups associated with a specific dynamical theory.
Let $G$ be a transformation group acting on the spatially periodic and
non-divergent function $\vec{v}\left( \vec{x}\text{, }t\right) $. $G$ is
said to be the symmetry group of the Navier-Stokes equations if, for all the 
$\vec{v}$ that are solutions of the Navier-Stokes equations, and $\forall
g\in G$, the function $g\vec{v}$ is also a solution. The known symmetries of
the Navier-Stokes equations are \cite{frish}:

\begin{enumerate}
\item Spatial translations $\left( t\text{, }\vec{x}\text{, }\vec{v}\right)
\rightarrow \left( t\text{, }\vec{x}+\vec{a}\text{, }\vec{v}\right) $ with $%
\vec{a}\in 
\mathbb{R}
^{3}$.

\item Temporal translations $\left( t\text{, }\vec{x}\text{, }\vec{v}\right)
\rightarrow \left( t+\tau \text{, }\vec{x}\text{, }\vec{v}\right) $ with $%
\tau \in 
\mathbb{R}
$.

\item Galilean transformations $\left( t\text{, }\vec{x}\text{, }\vec{v}%
\right) \rightarrow \left( t\text{, }\vec{x}+\vec{V}t\text{, }\vec{v}+\vec{V}%
\right) $ with $\vec{V}\in 
\mathbb{R}
^{3}$.

\item Parity $\left( t\text{, }\vec{x}\text{, }\vec{v}\right) \rightarrow
\left( t\text{, }-\vec{x}\text{, }-\vec{v}\right) $.

\item Rotation $\left( t\text{, }\vec{x}\text{, }\vec{v}\right) \rightarrow
\left( t\text{, }R\vec{x}\text{, }R\vec{v}\right) $ with $R\in SO(3,%
\mathbb{R}
)$.

\item Scaling $\left( t\text{, }\vec{x}\text{, }\vec{v}\right) \rightarrow
\left( \lambda ^{\left( 1-\alpha \right) }t\text{, }\lambda \vec{x}\text{, }%
\lambda ^{\alpha }\vec{v}\right) $ with $\lambda \in 
\mathbb{R}
_{+}$ and $\alpha \in 
\mathbb{R}
$.
\end{enumerate}

It is useful to observe that $5$ is valid in the limit where $l\rightarrow
\infty $ and that $6$ is valid for $\nu \rightarrow 0$ (that is, for large
Reynolds numbers). Furthermore, observe that the symmetry $\vec{v}%
\rightarrow -\vec{v}$ is inconsistent with Navier-Stokes equation except
when the non-linear term is negligible. Unlike the Euler equation, the
Navier-Stokes equation is not invariant under temporal inversion. This last
fact is a consequence of the emergence of dissipative phenomena. Finally,
notice that all symmetries except $6$ are nothing but macroscopic
consequences of fundamental symmetries of Newtons equations describing (in
classical approximation) the microscopic molecular motion of the fluid.
Before describing what happens when the Reynolds number increases in the
fluid, we define the concept of spontaneous symmetry breaking \cite{huang}.

It appears that macroscopic systems generally have a smaller degree of
symmetry at low rather than high temperatures. The manifested symmetry at
high temperatures is generally a property of the microscopic Hamiltonian of
the system. As such, it cannot cease to exist even when the symmetries
associated with the Hamiltonian appear to be violated. The question is,
where did the symmetry go? For example the microscopic Hamiltonian of a
ferromagnet is invariant under rotation. Lowering the temperature of the
system certainly does not change this fact. What actually changes is the
manner in which the symmetry manifests itself. It is natural to assume that
in the most perfect manifestation of symmetry the ground state should be
invariant under symmetry transformations. In many cases, the system has
several equivalent ground states that can be mapped into each other via
symmetry transformation. However, since the system can actually be in one
and only one of these ground states, the symmetry appears broken. When the
ground state of the system does not share the same symmetry of the
Hamiltonian, the symmetry is said to be spontaneously broken.

Returning to the issue of turbulence, it is experimentally observed that
when the Reynolds number increases, the symmetries permitted by the
equations of motion and the boundary conditions are subsequently broken.
However, for very high Reynolds numbers there appears to be a tendency to
restore the symmetries (in a statistical sense) far away from the
boundaries. That is to say, the symmetries are restored on average but not
for a single realization of the velocity field of the system. Such
turbulence is referred to as fully developed turbulence \cite{frish}. Fully
developed turbulence is turbulence that is free to develop without any
constraints. The only possible constraints are the boundaries, external
forces or viscosity. It is observed that the structures of a flow that
develops on scales comparable to the dimensions over which the fluid evolves
cannot be properly defined as "developed". For this reason, no real fluid -
even if has a high Reynolds number - can be "fully developed"\ on large
energy scales. On smaller scales however, turbulence will be fully developed
provided the viscosity does not play a direct role in the dynamics at such
scales.

The turbulent flow at sufficiently high Reynolds numbers is characterized by
an extremely irregular and disordered temporal variation of the velocity
field at each point. There is experimental evidence that the velocity field
of a turbulent flow is fractal in nature \cite{feder, gallavotti}. Turbulent
fluids seem to have fractal velocity fields in the sense that the increments
of the velocity field are proportional to the power $1/3$ of the increment
of space. In a turbulent flow at very large Reynolds numbers, the average
quadratic increment of the velocity between two points separated by distance 
$l$ scales approximately with the power $2/3$ of this distance \cite%
{gallavotti}%
\begin{equation}
\left\langle \left[ \delta v\left( l\right) \right] ^{2}\right\rangle \sim
l^{\frac{2}{3}}\text{.}
\end{equation}%
We will not proceed further with regard to turbulence and will introduce new
concepts and theories as required. From a mathematical point of view, the
central problem of turbulence theory is that of obtaining statistical
solutions of the Navier-Stokes equations. The standard techniques of fluid
mechanics are not sufficiently powerful to study turbulence. In the second
half of the last century a formal analogy has been found between the theory
of turbulence and quantum field theory. In both cases a system of
interacting fields is non-linear, in principle with an infinite number of
degrees of freedom. From here follows the similarity of the mathematical
apparatus used in both theories. For example, the method of Feynman diagrams 
\cite{sakurai} used to represent equations (transition amplitudes, etc.) may
also be applied to turbulence theory (tree diagrams, etc.). Our approach to
the study of turbulence is to treat it as an analytic statistical theory
(Kraichnan-Orszag) that relies on dimensional considerations and similarity 
\cite{panchev}.

\subsection{Incompressible Turbulence: Lighthill Process}

Very weak turbulence can be described by linearized dynamical equations \cite%
{monin}. Let us consider the case in which turbulence is relatively weak,
but not so much so that we are able to neglect the non-linear terms in the
dynamical equations. In this context, we consider as the main quadratic
effect the generation of sound by turbulence in a compressible medium. The
production of sound due to the self-interaction of turbulent vortices (this
is the main interaction that cause second order effects) happens only when
the compressibility of the medium is taken into consideration. As an example
concerning the generation of mass density fluctuations due to turbulent
phenomena, let us consider the wave equation of linear acoustics with a
source term due to turbulence \cite{landau-fluid}%
\begin{equation}
\left( \vec{\nabla}^{2}-\frac{1}{c_{s}^{2}}\frac{\partial ^{2}}{\partial
t^{2}}\right) \delta \rho \left( \vec{x}\text{, }t\right) =-\frac{\rho _{0}}{%
c_{s}^{2}}\frac{\partial ^{2}}{\partial x_{i}\partial x_{j}}v_{i}^{\prime
}v_{j}^{\prime }  \label{inter15}
\end{equation}%
Equation (\ref{inter15}) describes the so-called Lighthill process. This
phenomenon consists of the generation of acoustic noise due to the presence
of turbulent fluid flow. In other words, turbulence (or more accurately, the
fluctuations of turbulent velocity) generates sound \cite{landau-fluid}. We
wish to emphasize two points that are implicit in (\ref{inter15}):

\begin{enumerate}
\item Equation (\ref{inter15}) has been obtained by assuming we are dealing
with compressible turbulence. A fundamental difference between compressible
and incompressible turbulence is the following: for compressible turbulence,
variations in the velocity field imply variations in the local mass density
of the fluid. Fluctuations in the mass density imply local variations of
pressure that lead to the emission of sound waves. For incompressible
turbulence instead, changes in pressure do not produce changes in density.

\item In the source term of (\ref{inter15}) are present only the fluctuating
parts of the turbulent velocity field ( $\vec{v}=\vec{u}+\vec{v}^{\prime }$,
where $\vec{u}$ is the average velocity field and $\vec{v}^{\prime }$ is the
fluctuating velocity field). The quantity $\rho v_{i}^{\prime }v_{j}^{\prime
}$ is called the Reynolds strain. The laminar portion of the flow, if one
exists, does not play any role in generating turbulent fluctuations that
give rise to acoustic noise in the Lighthill process.
\end{enumerate}

For reasons of analytical complexity, we will not solve (\ref{inter15}). We
will instead study the simplified equation%
\begin{equation}
\left( \vec{\nabla}^{2}-\frac{1}{c_{s}^{2}}\frac{\partial ^{2}}{\partial
t^{2}}\right) \delta p\left( \vec{x}\text{, }t\right) =-\rho _{0}\frac{%
\partial ^{2}}{\partial x_{i}\partial x_{j}}v_{i}v_{j}  \label{inter16}
\end{equation}%
where the source term is due to incompressible turbulence, keeping the
hypothesis of compressibility in the medium where the sound waves propagate.
Equation (\ref{inter16}) has been derived by neglecting the effects related
to viscosity and thermal conductivity. Furthermore, it has been assumed that
the incompressible velocity fluctuations $\vec{v}^{\prime }$ are small
compared to the average speed of sound $c_{s}$. Equation (\ref{inter16}) can
then be viewed as describing the generation of sound by turbulence with a
small Mach number $M=U/c_{s}$ (where $U$ is the characteristic scale of the
velocity of the system) and not simply by decaying turbulence (that is,
turbulence approaching transition to laminar flow).

Equation (\ref{inter16}) leads to a number of important consequences. For
example the right hand side of the equation is a combination of second order
derivatives of the field $\vec{v}\left( \vec{x}\right) $. This means that in
absence of boundary conditions, the generation of sound waves from
turbulence is equivalent to the radiation emitted from a set of acoustic
quadrupoles (and not by the usual acoustic dipole sources). Therefore, it
follows that if there are no boundaries, at small Mach numbers the
turbulence does not represent an efficient source of sound. Thus, we assume
the "acoustic noise" is generated within a bounded region ("turbulent
region") of the fluid in which velocity fluctuations occur. We assume that
the medium surrounding this volume is at rest (this is a much more extended
region than the turbulent region and is called the "radiation
region"\textquotedblright ). In order to solve (\ref{inter16}) we apply
analytic statistical theory of turbulence \cite{monin}. To simplify
calculations we assume we are dealing with fully developed turbulence that
is homogenous, isotropic and stationary.

The turbulence is said to be homogenous if all quantities constructed with a
set of $n$ points $\vec{x}_{1}$,...,$\vec{x}_{n}$ (at instants $%
t_{1},...,t_{n}$) are invariant under any translation of this set. In
particular \cite{lesieur}%
\begin{equation}
\left\langle u_{\alpha _{1}}\left( \vec{x}_{1}\text{, }t_{1}\right) \text{%
.......}u_{\alpha _{n}}\left( \vec{x}_{n}\text{, }t_{n}\right) \right\rangle
=\left\langle u_{\alpha _{1}}\left( \vec{x}_{1}+\vec{y}\text{, }t_{1}\right) 
\text{.......}u_{\alpha _{n}}\left( \vec{x}_{n}+\vec{y}\text{, }t_{n}\right)
\right\rangle
\end{equation}%
where $\left\langle \text{ }\cdot \cdot \cdot \right\rangle $ is the usual
ensemble expectation value. The turbulence is said to be stationary if all
the average quantities involved in the $n$ instants ($t_{1},...,t_{n}$) are
invariant under any temporal translation. In particular \cite{lesieur} 
\begin{equation}
\left\langle u_{\alpha _{1}}\left( \vec{x}_{1}\text{, }t_{1}\right) \text{%
.......}u_{\alpha _{n}}\left( \vec{x}_{n}\text{, }t_{n}\right) \right\rangle
=\left\langle u_{\alpha _{1}}\left( \vec{x}_{1}\text{, }t_{1}+\tau \right) 
\text{.......}u_{\alpha _{n}}\left( \vec{x}_{n}\text{, }t_{n}+\tau \right)
\right\rangle
\end{equation}%
Finally, the homogenous turbulence is said to be isotropic if all the
average quantities concerning the set of $n$ points $\vec{x}_{1}$,...,$\vec{x%
}_{n}$ (at instants $t_{1}$,...,$t_{n}$) are invariant under any arbitrary
rotation.

It could be argued that there is no turbulent flows that are homogenous or
isotropic at large-scale. Isotropy and homogeneity can even be debatable at
small scales. Nevertheless, these hypotheses enable us to easily exploit
analytic statistical theories, thereby enormously simplifying the equations
of motion. Such theories are quite powerful in the sense that they allow to
deal with strong non-linearity when the deviation from the hypothesis of
non-Gaussianity is not large. The point of view adopted in this article is
that these (analytic-statistical) techniques describe in a satisfactory
manner the dynamics of three-dimensional turbulent flows at small scales.
The price of this simplification is the gap between the situation studied
theoretically and that which can be realized in practice. The homogeneity
hypothesis implies that turbulence is uniform in space and the concept of
stationarity can be described as homogeneity in time. Isotropy implies there
are no preferred directions in space. There cannot be any average velocity
in an isotropic field since that would immediately imply a preferred
direction. Turbulent isotropic and homogenous fields can be relatively
simple but they are unphysical. In actual velocity fields, energy arises
from some average gradient of pressure, temperature or mass, and therefore
these fields must be anisotropic. Moreover, these fields will be subject to
specific boundary conditions that imply they are necessarily inhomogeneous.

Despite these facts, we will consider in this work fully developed,
homogenous, isotropic and stationary turbulence. That said, we recast (\ref%
{inter16}) in frequency space%
\begin{equation}
\left( -\vec{k}^{2}+\frac{\omega ^{2}}{c_{s}^{2}}\right) \delta \tilde{p}%
\left( \vec{k}\text{, }\omega \right) =\rho _{0}k_{i}k_{j}\mathcal{\tilde{V}}%
_{ij}\left( \vec{k}\right)
\end{equation}%
with%
\begin{equation}
\mathcal{\tilde{V}}_{ij}(\vec{k})=\underset{V}{\int }d^{3}\vec{x}e^{i\vec{k}%
\cdot \vec{x}}v_{i}\left( \vec{x}\right) v_{j}\left( \vec{x}\right)
\end{equation}%
where $V$ is the volume occupied by the turbulent fluid and $\rho _{0}$ is
the average density. We obtain%
\begin{equation}
\delta \tilde{p}(\vec{k}\text{, }\omega )=\frac{\rho _{0}c_{s}^{2}}{\left(
\omega ^{2}-c_{s}^{2}\vec{k}^{2}\right) }k_{i}k_{j}\mathcal{\tilde{V}}_{ij}(%
\vec{k})
\end{equation}%
and thus the noise spectrum $\mathcal{\tilde{S}}_{p}(\vec{k}$, $\omega )$
associated to the pressure fluctuation becomes, in the frequency space,%
\begin{equation}
\mathcal{\tilde{S}}_{p}(\vec{k}\text{, }\omega )=\left\langle \delta \tilde{p%
}\left( \vec{k}\text{, }\omega \right) \delta \tilde{p}^{\ast }\left( \vec{k}%
\text{, }\omega \right) \right\rangle =\frac{\rho _{0}^{2}c_{s}^{4}}{\left(
\omega ^{2}-c_{s}^{2}\vec{k}^{2}\right) ^{2}}k_{i}k_{j}k_{l}k_{m}\left%
\langle \mathcal{\tilde{V}}_{ij}(\vec{k})\mathcal{\tilde{V}}_{lm}^{\ast }(%
\vec{k})\right\rangle \text{.}
\end{equation}%
Finally, the noise spectrum becomes%
\begin{equation}
\mathcal{\tilde{S}}_{p}(\vec{k}\text{, }\omega )=\frac{\rho _{0}^{2}c_{s}^{4}%
}{\left( \omega ^{2}-c_{s}^{2}\vec{k}^{2}\right) ^{2}}k_{i}k_{j}k_{l}k_{m}%
\mathcal{C}_{ijlm}(\vec{k})\text{,}  \label{doublep}
\end{equation}%
where%
\begin{equation}
\mathcal{C}_{ijlm}(\vec{k})=\left\langle \mathcal{\tilde{V}}_{ij}(\vec{k})%
\mathcal{\tilde{V}}_{lm}^{\ast }(\vec{k})\right\rangle \text{.}
\end{equation}%
Our task now is to compute the quantity $\mathcal{C}_{ijlm}(\vec{k})$, 
\begin{equation}
\mathcal{C}_{ijlm}(\vec{k})=\underset{V}{\int }d^{3}\vec{x}\underset{V}{\int 
}d^{3}\vec{x}^{\prime }e^{-i\vec{k}\cdot \left( \vec{x}^{\prime }-\vec{x}%
\right) }\mathcal{B}_{ij\text{, }lm}^{\left( 4\right) }\left( \vec{x}\text{, 
}\vec{x}^{\prime }\right)
\end{equation}%
with%
\begin{equation}
\mathcal{B}_{ij\text{, }lm}^{\left( 4\right) }\left( \vec{x}\text{, }\vec{x}%
^{\prime }\right) =\left\langle v_{i}\left( \vec{x}\right) v_{j}\left( \vec{x%
}\right) v_{l}\left( \vec{x}^{\prime }\right) v_{m}\left( \vec{x}^{\prime
}\right) \right\rangle \text{.}  \label{inter19}
\end{equation}%
The quantity $\mathcal{B}_{ij\text{, }lm}^{\left( 4\right) }\left( \vec{x}%
\text{, }\vec{x}^{\prime }\right) $ appearing in (\ref{inter19}) is known in
statistical fluid mechanics as the tensorial statistical moment \cite%
{panchev}. In the most general case, the stochastic moments of a stochastic
vectorial field represent tensors of order $k$ that have the form%
\begin{equation}
\mathcal{B}_{ij\text{.....}p}^{\left( k\right) }\left( \vec{x}_{1}\text{%
,...., }\vec{x}_{n}\right) \text{.}
\end{equation}%
Additional limiting conditions on such stochastic fields lead to new
symmetry properties of such tensors. The conditions of homogeneity and
isotropy are of particular interest since, in practice, they are the only
symmetries considered. However, even without these special conditions, the
expression of the statistical moment $\mathcal{B}_{ij...p}^{\left( k\right)
} $ cannot be arbitrary since it must satisfy special tensorial
transformations. The quantity $\mathcal{B}_{ij...p}^{\left( k\right) }\left( 
\vec{x}_{1},...,\vec{x}_{n}\right) $ is a tensor of forth order representing
a two-point statistical moment. As a working hypothesis we assume that the
fields $v_{i}\left( \vec{x}\right) $ are Gaussian, stochastic velocity
fields. Under such hypothesis we obtain (in analogy to what is obtained by
applying Wick's theorem in quantum field theory \cite{sakurai})%
\begin{eqnarray}
\left\langle v_{i}\left( \vec{x}\right) v_{j}\left( \vec{x}\right)
v_{l}\left( \vec{x}^{\prime }\right) v_{m}\left( \vec{x}^{\prime }\right)
\right\rangle &=&\left\langle v_{i}\left( \vec{x}\right) v_{j}\left( \vec{x}%
\right) \right\rangle \left\langle v_{l}\left( \vec{x}^{\prime }\right)
v_{m}\left( \vec{x}^{\prime }\right) \right\rangle +\left\langle v_{i}\left( 
\vec{x}\right) v_{l}\left( \vec{x}^{\prime }\right) \right\rangle
\left\langle v_{j}\left( \vec{x}\right) v_{m}\left( \vec{x}^{\prime }\right)
\right\rangle +  \label{correlation} \\
&&  \notag \\
&&+\left\langle v_{i}\left( \vec{x}\right) v_{m}\left( \vec{x}^{\prime
}\right) \right\rangle \left\langle v_{j}\left( \vec{x}\right) v_{l}\left( 
\vec{x}^{\prime }\right) \right\rangle \text{.}  \notag
\end{eqnarray}%
From turbulence theory it follows that \cite{leslie}%
\begin{equation}
\left\langle v_{i}\left( \vec{x}\right) v_{j}\left( \vec{x}\right)
\right\rangle =\frac{2}{3}\delta _{ij}\dint\limits_{0}^{+\infty }dk\mathcal{E%
}\left( k\right)  \label{two-point}
\end{equation}%
and%
\begin{equation}
\left\langle v_{i}\left( \vec{x}\right) v_{j}\left( \vec{x}^{\prime }\right)
\right\rangle =\int \frac{d^{3}\vec{k}}{\left( 2\pi \right) ^{3}}e^{-i\vec{k}%
\cdot \left( \vec{x}^{\prime }-\vec{x}\right) }\left( \delta _{ij}-\frac{%
k_{i}k_{j}}{\vec{k}^{2}}\right)  \label{other2}
\end{equation}%
where $\mathcal{E}\left( k\right) $ is the Kolmogorov energy spectrum, that
in the interval $k_{0}<<k<<k_{\nu }$ can be written as%
\begin{equation}
\mathcal{E}(k)=\mathcal{K}_{0}\varepsilon ^{\frac{2}{3}}k^{-\frac{5}{3}}%
\text{.}  \label{inter20}
\end{equation}%
The interval of validity of (\ref{inter20}) is characterized by $k_{0}\sim
2\pi /L$ and $k_{\nu }=\mathcal{R}^{3/4}/L=\left( \varepsilon /\nu
^{3}\right) ^{1/4}$ where $L$ is the linear dimension of the volume of the
turbulent fluid and $\mathcal{K}_{0}$ is the Kolmogorov constant. The
quantity $\varepsilon $ represents the total energy dissipated due to
viscous forces and is given by%
\begin{equation}
\varepsilon =\dint\limits_{0}^{+\infty }dk2\nu \mathcal{E}\left( k\right) 
\text{.}
\end{equation}%
The quantity $\mathcal{E}\left( k\right) $ in (\ref{inter20}) represents the
Kolmogorov energy spectrum in the so-called inertial range. This is
permitted since we are considering a turbulence problem and such turbulence
is characterized by inertial modes ($k_{0}<<k<<k_{\nu }$) while the motion
of dissipative modes ($k_{\nu }<<k<<k_{\nu }^{\prime }$, $k_{\nu }^{\prime }$
is the viscous scale) is always laminar (the function of dissipative modes
is to absorb energy from inertial modes and dissipate it). Therefore, using (%
\ref{correlation}), (\ref{two-point}), (\ref{other2}) and (\ref{inter20})
the quantity $\mathcal{C}_{ijlm}(\vec{k})$ becomes%
\begin{eqnarray}
\mathcal{C}_{ijlm}(\vec{k}) &=&\frac{4}{9}\delta _{ij}\delta _{lm}\left(
\dint\limits_{k_{0}}^{k_{v}}dk\mathcal{E}\left( k\right) \right) ^{2}%
\underset{V}{\int }d^{3}\vec{x}\underset{V}{\int }d^{3}\vec{x}^{\prime }e^{-i%
\vec{k}\cdot \left( \vec{x}^{\prime }-\vec{x}\right) }+  \label{inter21} \\
&&  \notag \\
&&+\frac{4}{9}\left( \delta _{il}\delta _{jm}+\delta _{im}\delta
_{jl}\right) \underset{V}{\int }d^{3}\vec{x}\underset{V}{\int }d^{3}\vec{x}%
^{\prime }e^{-i\vec{k}\cdot \left( \vec{x}^{\prime }-\vec{x}\right) }\left(
\dint\limits_{k_{0}}^{k_{v}}dk\mathcal{E}\left( k\right) \frac{\sin \left(
kl\right) }{kl}\right) ^{2}  \notag
\end{eqnarray}%
with $l=\left\vert \vec{x}^{\prime }-\vec{x}\right\vert $. For economy of
notation, we define%
\begin{equation}
\mathcal{I}_{k_{0}k_{v}}^{\left( 1\right) }(\vec{k})=\underset{V}{\int }d^{3}%
\vec{x}\underset{V}{\int }d^{3}\vec{x}^{\prime }e^{-i\vec{k}\cdot \left( 
\vec{x}^{\prime }-\vec{x}\right) }\left( \dint\limits_{k_{0}}^{k_{v}}dk%
\mathcal{E}\left( k\right) \right) ^{2}  \label{inter22}
\end{equation}%
and%
\begin{equation}
\mathcal{I}_{k_{0}k_{v}}^{\left( 2\right) }(\vec{k})=\underset{V}{\int }d^{3}%
\vec{x}\underset{V}{\int }d^{3}\vec{x}^{\prime }e^{-i\vec{k}\cdot \left( 
\vec{x}^{\prime }-\vec{x}\right) }\left( \dint\limits_{k_{0}}^{k_{v}}dk%
\mathcal{E}\left( k\right) \frac{\sin \left( kl\right) }{kl}\right) ^{2}%
\text{.}  \label{inter23}
\end{equation}%
In terms of (\ref{inter22}) and (\ref{inter23}) the quantity $\mathcal{C}%
_{ijlm}(\vec{k})$ in (\ref{inter21}) can be written as%
\begin{equation}
\mathcal{C}_{ijlm}(\vec{k})=\frac{4}{9}\delta _{ij}\delta _{lm}\mathcal{I}%
_{k_{0}k_{v}}^{\left( 1\right) }(\vec{k})+\frac{4}{9}\left( \delta
_{il}\delta _{jm}+\delta _{im}\delta _{jl}\right) \mathcal{I}%
_{k_{0}k_{v}}^{\left( 2\right) }(\vec{k})\text{,}
\end{equation}%
such that%
\begin{equation}
k_{i}k_{j}k_{l}k_{m}\mathcal{C}_{ijlm}(\vec{k})=\frac{4}{9}\vec{k}^{4}\left( 
\mathcal{I}_{k_{0}k_{v}}^{\left( 1\right) }(\vec{k})+2\mathcal{I}%
_{k_{0}k_{v}}^{\left( 2\right) }(\vec{k})\right) \text{.}  \label{inter24}
\end{equation}%
Placing (\ref{inter24}) in (\ref{doublep}), the noise spectrum $\mathcal{%
\tilde{S}}_{p}(\vec{k}$, $\omega )$ associated to the pressure fluctuation
becomes%
\begin{equation}
\mathcal{\tilde{S}}_{p}(\vec{k},\omega )=\frac{4}{9}\rho
_{0}^{2}c_{s}^{4}\left( \mathcal{I}_{k_{0}k_{v}}^{\left( 1\right) }(\vec{k}%
)+2\mathcal{I}_{k_{0}k_{v}}^{\left( 2\right) }(\vec{k})\right) \frac{\vec{k}%
^{4}}{\left( \omega ^{2}-c_{s}^{2}\vec{k}^{2}\right) ^{2}}\text{.}
\label{inter26}
\end{equation}%
Performing a Fourier transform of $\left( \mathcal{S}_{a}\right) _{ij}\left( 
\vec{x}_{1}\text{, }\vec{x}_{2}\text{; }\omega \right) $ in (\ref{inter4})
in the $\vec{x}$ variable, we obtain 
\begin{equation}
\mathcal{\tilde{S}}_{\vec{a}}(\vec{k},\omega )\equiv \left\langle \tilde{a}%
_{i}\left( \vec{k}\text{, }\omega \right) \tilde{a}_{j}^{\ast }\left( \vec{k}%
\text{, }\omega \right) \right\rangle =\left( \frac{4\pi G}{c_{s}^{2}}%
\right) ^{2}\frac{k_{i}k_{j}}{k^{4}}\mathcal{\tilde{S}}_{p}(\vec{k},\omega )%
\text{.}  \label{inter25}
\end{equation}%
The quantity $\mathcal{\tilde{S}}_{\vec{a}}(\vec{k},\omega )$ in (\ref%
{inter25}) represents the correlation function associated with the
fluctuation of acceleration in the Fourier space in terms of the spectrum of
pressure described in the frequency space. Substituting (\ref{inter26}) into
(\ref{inter25}) we find%
\begin{equation}
\mathcal{\tilde{S}}_{\vec{a}}(\vec{k},\omega )=\frac{4}{9}\left( 4\pi \rho
_{0}G\right) ^{2}\left( \mathcal{I}_{k_{0}k_{v}}^{\left( 1\right) }(\vec{k}%
)+2\mathcal{I}_{k_{0}k_{v}}^{\left( 2\right) }(\vec{k})\right) \frac{%
k_{i}k_{j}}{\left( \omega ^{2}-c_{s}^{2}\vec{k}^{2}\right) ^{2}}\text{.}
\end{equation}%
At this point all that remains is to evaluate the integrals $\mathcal{I}%
^{\left( 1\right) }(\vec{k})$ and $\mathcal{I}^{\left( 2\right) }(\vec{k})$.
For $\mathcal{I}^{\left( 1\right) }(\vec{k})$ we find the explicit
functional form%
\begin{equation}
\mathcal{I}_{k_{0}k_{v}}^{\left( 1\right) }(\vec{k})=\frac{9}{4}\mathcal{K}%
_{0}^{2}\varepsilon ^{\frac{4}{3}}\left( k_{v}^{-\frac{2}{3}}-k_{0}^{-\frac{2%
}{3}}\right) ^{2}\frac{16\pi ^{2}}{k^{6}}\left[ \sin \left( kL\right)
-kL\cos \left( kL\right) \right] ^{2}\text{.}
\end{equation}%
Concerning $\mathcal{I}^{\left( 2\right) }(\vec{k})$\textbf{, }we notice it
can be recast in the following form,%
\begin{equation}
\mathcal{I}_{k_{0}k_{v}}^{\left( 2\right) }(\vec{k})=\frac{4\pi \left( \frac{%
4\pi }{3}L^{3}\right) \mathcal{K}_{0}^{2}\varepsilon ^{\frac{4}{3}}}{k}%
\dint\limits_{0}^{L}dll^{\frac{7}{3}}\sin \left( kl\right) \left(
\dint\limits_{k_{0}l}^{k_{v}l}d\alpha \frac{\sin \alpha }{\alpha ^{\frac{8}{3%
}}}\right) ^{2}\text{.}  \label{q}
\end{equation}%
The integral in $\alpha $ can be expressed exactly in terms of incomplete
Gamma functions $\Gamma \left( a\text{, }z\right) $ \cite{morse},%
\begin{equation}
\Gamma \left( a\text{, }z\right) \overset{\text{def}}{=}\int_{z}^{\infty
}t^{a-1}\exp \left( -t\right) dt\text{.}
\end{equation}%
The remaining expression in (\ref{q}) can be numerically integrated for a
given choice of the parameters in play. The numerical values of the
parameters used in our calculations is given in Table 1. The theoretical
value of the Kolmogorov constant $\mathcal{K}_{0}$ is borrowed from
Reference \cite{push}. 
\begin{equation*}
\underset{\text{Table 1}}{%
\begin{tabular}{|c|c|c|}
\hline
Parameter & Symbol & Value $\left[ \text{MKSA-units}\right] $ \\ \hline
linear scale of turbulent region & $L$ & $150$ \\ \hline
Newton's constant & $G$ & $6.67\times 10^{-11}$ \\ \hline
kinematic viscosity of air at $T=15^{o}C$ & $\nu _{0}$ & $1.8\times 10^{-5}$
\\ \hline
air density & $\rho _{0}$ & $1.3$ \\ \hline
Reynolds number & $R$ & $3200$ \\ \hline
Kolmogorov constant & $\mathcal{K}_{0}$ & $9.85$ \\ \hline
\end{tabular}%
}
\end{equation*}%
Our numerical estimate leads to the square root of the strain amplitude of
the atmospheric Newtonian noise\textbf{\ }$\tilde{h}_{ANN}\left( f\right) $%
\textbf{\ }given by,%
\begin{equation}
\tilde{h}_{ANN}\left( f\right) \overset{\text{def}}{=}\sqrt{\left\langle
\left\vert \tilde{h}_{ANN}\left( f\right) \right\vert ^{2}\right\rangle }=%
\frac{1}{\left( 2\pi \right) ^{2}L}\frac{1}{f^{2}}\sqrt{\mathcal{\tilde{S}}_{%
\vec{a}}\left( \vec{k}\text{, }\omega \right) }\text{.}
\end{equation}%
Recall that the sensitivity of the Virgo detector is quantified in terms of
the square root of the strain amplitude $\tilde{h}_{rss}\left( f\right) $%
\textbf{. }For instance, at $f\simeq 360$Hz the best sensitivity at $50\%$
of efficiency is $\left[ \tilde{h}_{rss}^{50\%}\left( f\right) \right]
_{f\simeq 360Hz}\approx 1.1\times 10^{-20}/\sqrt{\text{Hz}}$ \cite{acer}.%
\textbf{\ }In our work, we have compared Virgo's $\tilde{h}_{rss}\left(
f\right) $ to $\tilde{h}_{ANN}\left( f\right) $ in the frequency range $f\in %
\left[ 4\text{, }10\right] $\textbf{. }We have,%
\begin{equation}
\left[ \frac{\tilde{h}_{ANN}\left( f\right) }{\tilde{h}_{rss}\left( f\right) 
}\right] _{f\in \left[ 4\text{, }10\right] \text{Hz}}\approx \frac{10^{-23}}{%
10^{-20}}=10^{-3}\ll 1\text{.}
\end{equation}%
It turns out that the effect of acoustic noise generated in the Lighthill
process considered here is at least three orders of magnitude below the
sensitivity of the VIRGO interferometer.

\section{Conclusions}

In this article, we presented a theoretical estimate of the atmospheric
Newtonian noise generated by fluctuations of atmospheric mass densities due
to acoustic and turbulent phenomena and we judge the relevance of such noise
in the laser-interferometric detection of gravitational waves. First, we
considered the gravitational coupling of interferometer test-masses to
fluctuations of atmospheric density due to the propagation of sound waves in
a semispace occupied by an ideal fluid delimited by an infinitely rigid
plane. We presented an analytical expression of the spectrum of acceleration
fluctuations $\mathcal{S}_{a}\left( \vec{r}_{1}\text{, }\vec{r}_{2}\text{; }%
\omega \right) $ of the test-masses of the interferometer in terms of the
experimentally determinable spectrum of pressure fluctuations $\mathcal{S}%
_{p}\left( \vec{r}_{1}\text{, }\vec{r}_{2}\text{; }\omega \right) $. We do
not have direct measurements of $\mathcal{S}_{p}$ for the VIRGO\ detector.
However, values extracted from the literature lead to conclude that the
effect would be at least two orders of magnitude below the sensitivity
curve. Second, we considered the gravitational coupling of interferometer
test-masses to fluctuations of atmospheric density due to the propagation of
sound waves generated in a turbulent Lighthill process. We presented an
analytical expression, in the Fourier space, of the spectrum of acceleration
fluctuations $\mathcal{\tilde{S}}_{a}\left( \vec{k}_{1}\text{, }\vec{k}_{2}%
\text{; }\omega \right) $\ of the test-masses of the interferometer. We
estimated that the acoustic noise generated in the Lighthill process is
three orders of magnitude below the sensitivity curve of the VIRGO
interferometer.

\begin{acknowledgments}
C. C. thanks Prof. Carlo Bradaschia and especially Dr. Giancarlo Cella for
their guidance at the first stage of this work carried on at the University
of Pisa, Italy. Moreover, he thanks Giuseppe Aiezza, Salvatore Aiezza and
Joseph Cafaro for technical assistance. Finally, we thank the Referee for
very challenging comments.
\end{acknowledgments}


\begin{thebibliography}{99}
\bibitem{MTW} C. W. Misner, K. S. Thorne and J. Wheeler, "\emph{Gravitation}%
", W. H. Freeman \& Company- San Francisco (1973).

\bibitem{HT} R. A. Hulse and J. H. Taylor, \textit{Astrophys. Jour.} \textbf{%
195}, L51 (1975).

\bibitem{barone} M. Barone et. \textit{al.}, "\emph{Experimental Physics of
Gravitational Waves}", World Scientific (2000); A. Giazzotto, "\emph{%
Interferometric detection of gravitational waves}", Phys. Rep. \textbf{182},
365-424 (1989).

\bibitem{corda} Ch. Corda, "\emph{Interferometric detection of gravitational
waves: the definitive test for General Relativity}", arXiv:gr-qc/0905.2502
(2009).

\bibitem{nojiri} S. Nojiri and S. D. Odintsov, "\emph{Introduction to
Modified Gravity and Gravitational Alternatives for Dark Energy}", Int. J.
Geom. Meth. Mod. Phys. \textbf{4}, 115 (2007).

\bibitem{virgo} C. Bradaschia et. \textit{al}., Nuclear Instruments and
Methods in Physics Research \textbf{A289}, 518-525 (1990); F. Acernese et. 
\textit{al}., "\emph{Status of VIRGO}", Class. Quant. Grav. \textbf{22},
869-880 (2005).

\bibitem{weiss} R. Weiss, "\emph{Quarterly Progress Report of the Research
Laboratory of Electronics of the Massachusetts Institute of Technology}", 
\textbf{105}, 54 (1972).

\bibitem{saulson} P. R. Saulson, "\emph{Terrestrial and gravitational noise
on gravitational wave antenna}", Phys. Rev \textbf{D30}, 732-736 (1984).

\bibitem{spero} P. Spero, in \textit{Science Underground}, Proceedings of
the Los Alomos Conference, edited by M. M. Nieto et. al. (AIP, New York,
1983).

\bibitem{thorne} S. A. Hughes and K. S. Thorne, "\emph{Seismic
gravity-gradient noise in interferometric gravitational-wave detectors}",
Phys. Rev. \textbf{D58}, 122002 (1998); K. S. Thorne and C. J. Winstein, "%
\emph{Human gravity-gradient noise in interferometric gravitational-wave
detectors}", Phys. Rev. \textbf{D60}, 082001 (1999).

\bibitem{creighton} T. Creighton, "\emph{Tumbleweeds and Airborne
Gravitational Noise Sources for LIGO}", arXiv: gr-qc/0007050 (2000).

\bibitem{cafaro} C. Cafaro, "\emph{Stima Teorica del Rumore Newtoniano
Atmosferico in Rivelatori Interferometrici di Onde Gravitazionali}", MS{}
Thesis, University of Pisa, Italy (2002); available at \emph{%
www.virgo.infn.it/theses}.

\bibitem{landau-fluid} L. D. Landau and E. M. Lifshitz, "\emph{Fluid
Mechanics}" (Course of Theoretical Physics, vol. 6), Butterworth-Heinemann
(1987).

\bibitem{pp} E. S. Posmentier, J. Geophys. Res. \textbf{79}, 1755 (1974).

\bibitem{frish} U. Frish, "\emph{Turbulence}", Cambridge University Press
(1995).

\bibitem{huang} K. Huang, "\emph{Statistical Mechanics}", John Wiley \&
Sons, Inc. (1987).

\bibitem{feder} J. Feder, "\emph{Fractals}", Plenum Press (1989).

\bibitem{gallavotti} G. Gallavotti, "\emph{Ipotesi per una Introduzione alla
Meccanica dei Fluidi}", Quaderni del Consiglio Nazionale delle Ricerche 
\textbf{52}, Gruppo di Fisica Matematica (1996).

\bibitem{sakurai} J. J. Sakurai, "\emph{Advanced Quantum Mechanics}",
Addison-Wesley Series in Advanced Physics (1967).

\bibitem{panchev} S. Panchev, "\emph{Random Functions and Turbulence}",
Pergamon Press (1971).

\bibitem{monin} A.\ S. Monin and A. M. Yaglom, "\emph{Statistical Fluid
Mechanics: Mechanics of Turbulence, Volume II}" (vol. 2), MIT Press,
Cambridge (1975).

\bibitem{lesieur} M. Lesieur, "\emph{Turbulence in Fluids}", Kluwer Academic
Publishers (1997).

\bibitem{leslie} D. C. Leslie, "\emph{Developments in the Theory of
Turbulence}", Clarendon Press, Oxford (1972).

\bibitem{morse} P. M. Morse and H. Feshbach, "\emph{Methods of Theoretical
Physics}", McGraw-Hill (1953).

\bibitem{push} A. N. Pushkarev, "\emph{On the Kolmogorov and Frozen
Turbulence in Numerical Simulation of Capillary Waves}", Eur. J. Mech. 
\textbf{B18}, 345-351 (1999).

\bibitem{acer} Acernese et. \textit{al}., "\emph{Gravitational wave burst
search in the Virgo C7 data}", arXiv:0812.4870 (2008).
\end{thebibliography}
\end{document}